%% file: main.tex
\documentclass[runningheads]{llncs}

\input{preamble/packages}

\input{preamble/authors}

\begin{document}

\title{Efficient Optimization of Hierarchical Identifiers for Generative Recommendation}
\titlerunning{Efficient Optimization of Hierarchical Identifiers}
%
\maketitle             
\begingroup
\renewcommand\thefootnote{}
\footnotetext{* Equal contribution. \quad $\dagger$ Corresponding author.}
\endgroup

\begin{abstract}
SEATER is a generative retrieval model that improves recommendation inference efficiency and retrieval quality by utilizing balanced tree-structured item identifiers and contrastive training objectives. We reproduce and validate SEATER’s reported improvements in retrieval quality over strong baselines across all datasets from the original work, and extend the evaluation to Yambda, a large-scale music recommendation dataset. Our experiments verify SEATER’s strong performance, but show that its tree construction step during training becomes a major bottleneck as the number of items grows. To address this, we implement and evaluate two alternative construction algorithms: a greedy method optimized for minimal build time, and a hybrid method that combines greedy clustering at high levels with more precise grouping at lower levels. The greedy method reduces tree construction time to less than 2\% of the original with only a minor drop in quality on the dataset with the largest item collection. The hybrid method achieves retrieval quality on par with the original, and even improves on the largest dataset, while cutting construction time to just 5–8\%. All data and code are publicly available for full reproducibility at \url{https://github.com/joshrosie/re-seater}.

\keywords{Generative Recommendations \and Hierarchical Identifiers \and Reproducibility \and Scalability \and Recommender Systems}
\end{abstract}

\input{sections/sec-intro}
\input{sections/sec-rel}
\input{sections/sec-method}
\input{sections/sec-exp}

\input{sections/sec-results}
\input{sections/sec-con}




\bibliographystyle{splncs04}
\bibliography{references}

\end{document}

%% file: preamble/packages.tex
\usepackage[T1]{fontenc}
\usepackage[ruled,vlined]{algorithm2e}
\usepackage{orcidlink}
\usepackage{graphicx}
\usepackage{color}
\usepackage{amsmath}    
\usepackage{xcolor}
\usepackage{comment}
\usepackage{amsfonts}
\usepackage{booktabs}
\usepackage{multirow}
\usepackage{adjustbox}
\usepackage{verbatim}
\usepackage{subcaption}
\usepackage[table,xcdraw]{xcolor}

\usepackage{hyperref}
\hypersetup{
  colorlinks   = true,    
  urlcolor     = blue,    
  linkcolor    = blue,    
  citecolor    = red      
}

\usepackage{todonotes}
\usepackage{enumitem}

%% file: preamble/authors.tex
\author{
Federica Valeau \orcidlink{0009-0002-4959-0999}\textsuperscript{*$\dagger$} \and
Odysseas Boufalis \orcidlink{0009-0002-8703-466X}\textsuperscript{*} \and
Polytimi Gkotsi \orcidlink{0009-0007-0107-9784}\textsuperscript{*} \and
Joshua Rosenthal \orcidlink{0009-0005-9556-2982}\textsuperscript{*} \and
David Vos \orcidlink{0009-0003-8925-1585}
}

\authorrunning{Valeau et al.}

\institute{University of Amsterdam\\
\email{\{federica.valeau, odyssefs.boufalis, polytimi.gkotsi, joshua.rosenthal\}@student.uva.nl, d.j.a.vos@uva.nl}}

%% file: sections/sec-intro.tex
\section{Introduction}

During the retrieval phase of modern two-stage recommenders, efficiency and retrieval quality must be carefully balanced. Efficiency is essential because recommenders often operate at web scale, where millions of items must be searched in real time under strict latency constraints. At the same time, high retrieval quality is crucial, since the later ranking stage cannot easily correct errors made during retrieval.

Traditional retrieval models address this challenge through dual-encoder architectures, which encode users and items into a shared representation space. Retrieval is then performed through (approximate) nearest neighbor algorithms over these embeddings. This approach relies on the simplifying assumption that user-item relevance can be captured by an inner product in the embedding space. Previous research argues that this assumption often fails, leading to degraded recommendation quality \cite{comirec}.

Generative retrieval models have recently emerged as a promising alternative. Framing retrieval as a sequence generation task offers greater flexibility in how users and items are represented and matched. However, these models typically face two important limitations: (i) inference efficiency at a large scale, and (ii) retrieval quality degradation due to insufficient exploitation of the compositional semantics of items \cite{10.1145/2959100.2959190}.

To address these challenges of inference efficiency and retrieval quality, Si et al. introduced SEATER \cite{seater}. SEATER improves inference efficiency by mapping items to balanced, tree-structured identifiers, while preserving retrieval quality through contrastive training objectives that enforce semantic consistency across the tree. Their results show that their identifier design can both improve inference latency and retrieval quality. 

While SEATER improves inference efficiency, its tree-construction algorithm introduces a substantial training bottleneck, especially as the item catalog grows. Our experimental results confirm this, with tree construction time taking up to 19.4\% of total training time on the largest dataset (\autoref{tab:dataset_model_comparison_final_times}). In large-scale, frequently updated recommendation settings, this step becomes a key limitation, as tree construction must often be repeated when new items or user interactions appear. Generative retrieval models are particularly sensitive to this issue, since the retrieval index is encoded directly in model parameters, making fast retraining essential.

Given the importance of training efficiency in real-world scenarios, it is critical that improved inference speed and retrieval quality do not negatively impact tree-construction time. We first reproduce SEATER’s results to validate its reported gains in retrieval quality over established baselines. We extend the evaluation to a larger dataset, Yambda, a large-scale music recommendation dataset, allowing us to assess SEATER’s scalability. This leads to our first research question:

\begin{enumerate}[label=(RQ\arabic*),leftmargin=*,ref={RQ\arabic*}]
    \item Can the reported performance gains of SEATER over baselines be replicated, and generalized to the large-scale Yambda dataset? 
    \label{rq1}
\end{enumerate}

We then turn to the training bottleneck introduced by the tree construction algorithm and analyze its impact as the number of items increases. To mitigate this, we propose a modification to the procedure that replaces the global capacity-constrained optimization with a greedy assignment mechanism, in which items are sequentially assigned to the nearest centroid until cluster capacity is reached:

\begin{enumerate}[label=(RQ\arabic*),resume,leftmargin=*,ref={RQ\arabic*}]
    \item How does greedy clustering during tree construction impact the training efficiency and retrieval quality of SEATER?
    \label{rq2}
\end{enumerate}

Because early tree levels group large numbers of items, precise clustering at these stages may not be necessary. We hypothesize that combining greedy clustering in early steps with complete clustering in later steps can offer a balance between scalability and accuracy. This motivates our third question:

\begin{enumerate}[label=(RQ\arabic*),resume,leftmargin=*,ref={RQ\arabic*}]
    \item Can greedy clustering in early steps and complete clustering in later steps allow us to balance the training efficiency and retrieval quality of SEATER?
    \label{rq3}
\end{enumerate}

We verify that SEATER’s tree-construction algorithm indeed improves retrieval quality over strong baselines, including the new Yambda dataset (RQ1). We then demonstrate that a greedy clustering method massively reduces tree construction time, to less than 2\% of the original on the largest dataset, Yambda, while causing only a minor drop in retrieval quality (RQ2). Finally, we show that a hybrid method, combining greedy clustering at higher tree levels with full constrained clustering at lower levels, achieves retrieval quality on par with or even exceeding the original, while cutting construction time to just 5-8\% across datasets (RQ3). We publish all of our code for full reproducibility.\footnote{\url{https://github.com/joshrosie/re-seater}}

%% file: sections/sec-rel.tex
\section{Related work}

\subsubsection{Retrieval for recommender systems}
Traditional retrieval models for recommendations use dual-encoder architectures to represent users and items as embeddings. Efficient (approximate) nearest-neighbor algorithms then perform matching between user and item representations \cite{10.1145/2959100.2959190}. Models such as GRU4Rec \cite{gru4rec}, SASRec \cite{sasrec}, and BERT4Rec \cite{bert4rec} predict the next item from user interaction histories, typically treating item IDs as discrete tokens and using recurrent or transformer-based architectures to capture temporal patterns. Alternatively, tree-based methods like TDM \cite{tdm18} and JTM \cite{jtm20} improve retrieval efficiency by hierarchically organizing items for faster candidate generation. SEATER builds on these tree-based approaches by learning to optimize item identifiers from an indexing perspective \cite{seater}.

\vspace{-12px}
\subsubsection{Generative retrieval}
Generative retrieval frames retrieval as sequence generation over discrete identifiers, and was first explored in document search with models such as DSI and NCI \cite{10.1145/3476415.3476428,DBLP:conf/nips/Tay00NBM000GSCM22,10.1145/3626772.3657697}. In recommendation, TIGER \cite{rajput2023recommender} adapts this paradigm by learning semantic identifiers for items via vector quantization and training an encoder-decoder model to generate the corresponding code sequences. While effective, TIGER’s identifiers are not explicitly constrained to follow a structured hierarchy and therefore do not optimize structural properties during training. SEATER extends this paradigm by \emph{learning balanced, tree-structured identifiers} and introducing contrastive objectives that align the semantic and structural properties of the learned tree. Decoding is performed through constrained beam search restricted to valid tree paths, making the retrieval process explicitly index-aware.

\vspace{-1px}
\subsubsection{Document identifier construction}
A key component of generative retrieval is the design of discrete item or document identifiers. Early work in this area used vector quantization and discrete code learning to represent semantic content compactly \cite{pq,semantic-hashing,vqvae}. In recommendation, models such as TIGER \cite{rajput2023recommender} adopt this idea by mapping item embeddings to sequences of quantized codes, allowing an encoder-decoder model to autoregressively generate identifiers. While this captures semantic similarity, these identifiers are not explicitly optimized for structural consistency, resulting in uneven semantic granularity across items.

SEATER takes a different approach by \emph{explicitly constructing balanced, hierarchical identifiers}. Items are organized into a $k$-ary tree through constrained clustering, ensuring equal-length identifiers where tokens represent increasingly fine-grained semantic partitions. By coupling this construction with contrastive learning to align item semantics with the tree structure \cite{seater}, SEATER produces identifiers that combine semantic coherence with structural regularity, yielding faster inference and improved retrieval quality.

%% file: sections/sec-method.tex
\section{Methodology}

We first summarize the SEATER model \cite{seater}, which we reproduce as the basis for our study. We then present our proposed modifications to its tree construction algorithm to improve training efficiency while maintaining inference efficiency and retrieval quality.

\subsection{SEATER}

Given a user's interaction history, SEATER generates an item identifier: a sequence of tokens that uniquely represents the item to be retrieved. The model uses a lightweight transformer encoder-decoder architecture to perform this task. Each item in the catalog is assigned a unique identifier composed of multiple tokens, corresponding to nodes in a pre-constructed balanced $k$-ary tree. During retrieval, SEATER generates these identifiers using constrained beam search, ensuring that decoding follows valid token paths within the tree.

\subsubsection{Item identifiers}

Each token corresponds to a node in a balanced $k$-ary tree built before training. The original SEATER implementation constructs this tree using constrained $k$-means clustering over item embeddings obtained from a pretrained recommendation model such as SASRec. Items are recursively clustered into $k$ groups per level until each leaf group contains fewer than $k$ items. This hierarchical process groups semantically similar items together, resulting in similar identifiers for related items.

The balanced tree structure reduces search complexity from $O(bN)$ in naive decoding to $O(bk \log_k N)$, where $b$ is the beam width and $N$ the number of items. This leads to substantial inference-time efficiency gains over prior generative retrieval models \cite{seater}, as summarized in Table~\ref{tab:inference-comparison}.

\begin{table}[t!]
\centering
\begin{tabular}{lcc}
\hline
\textbf{Model} & \textbf{Inference Complexity} & \textbf{Identifier Size} \\
\hline
TDM \cite{tdm18} & $\mathcal{O}(b \log_{2} N)$ & $\mathcal{O}(Nd)$ \\
DSI \cite{DBLP:conf/nips/Tay00NBM000GSCM22} & $\mathcal{O}(bkL)$ & $\mathcal{O}(kd)$ \\
NCI \cite{NCI} & $\mathcal{O}(bkL)$ & $\mathcal{O}(kLd)$ \\
SEATER & $\mathcal{O}(bk \log_{k} N)$ & $\mathcal{O}(Nd)$ \\
\hline
\end{tabular}
\caption{Comparison of inference complexity and identifier size of retrieval models that use tree-based item identifiers.
$N$: number of items; $k$: branching factor; $b$: beam width; $d$: embedding dimension; $L$: identifier length (typically grows with $\log_k N$).}
\label{tab:inference-comparison}
\end{table}

\subsubsection{Training objectives}

To align the learned identifiers with SEATER’s retrieval objective, the model is optimized using a multi-task loss composed of three components:
\[
\mathcal{L} = \mathcal{L}_{\text{gen}} + \lambda_a \mathcal{L}_{\text{ali}} + \lambda_r \mathcal{L}_{\text{rank}},
\]
where $\lambda_a$ and $\lambda_r$ are weighting coefficients.

\begin{itemize}
    \item \textbf{Generation loss} ($\mathcal{L}_{\text{gen}}$): cross-entropy loss for autoregressive prediction of the ground-truth item identifier.
    \item \textbf{Alignment loss} ($\mathcal{L}_{\text{ali}}$): InfoNCE contrastive loss aligning each child node’s embedding with its parent, enforcing tree-level semantic consistency.
    \item \textbf{Ranking loss} ($\mathcal{L}_{\text{rank}}$): triplet-based contrastive loss encouraging items that share longer prefixes (i.e., are semantically closer) to rank higher.
\end{itemize}

These objectives jointly optimize SEATER to learn identifiers that are semantically meaningful and structurally coherent, leading to strong retrieval performance. However, the computational cost of constructing the item identifier tree remains a major bottleneck, motivating our investigation into more scalable alternatives.

\subsection{Tree construction algorithms}
\label{sec:tree-construction}

A core component of SEATER is its hierarchical tree construction, which maps items to balanced, semantically coherent identifiers. Each level of the tree is built using a capacity-constrained $k$-means algorithm that enforces equal cluster sizes to maintain balance. While this constraint improves inference efficiency and retrieval quality, it introduces a significant training bottleneck.

Given $N$ items and $k$ clusters, each constrained $k$-means operation involves solving a capacity-constrained assignment problem with worst-case complexity
\[
O\!\Bigl((N^{3}k + N^{2}k^{2} + N k^{3}) \log (N + k)\Bigr),
\]
which scales poorly as $N$ increases \cite{Levy-Kramer_k-means-constrained_2018}.
Moreover, because the capacity-constrained assignment step couples all item-cluster decisions, it is difficult to parallelize, limiting scalability on modern GPU hardware. As shown in \autoref{fig:tree-time-items}, this constrained $k$-means solution towards a balanced tree construction restricts SEATER’s practicality for large datasets or frequent retraining.

\subsubsection{Greedy clustering}
Our first proposed change replaces the global capacity-constrained optimization with a local greedy heuristic. Specifically, we first compute centroids using standard $k$-means, then sequentially assign items to the nearest available cluster subject to capacity limits, with $N/k$ items per cluster. This approach maintains balanced cluster sizes while reducing computational complexity to $O(Nk)$. We implement this method using Faiss for approximate nearest-neighbor search with GPU acceleration, enabling efficient parallel distance computations and further improving scalability. The result is an efficiently parallelizable, linear-time, balanced clustering method that preserves SEATER’s identifier semantics.

\vspace{-12px}

\subsubsection{Hybrid clustering}
While the greedy assignment greatly improves scalability, it may lead to suboptimal clusters due to its local, sequential nature. To mitigate this, we introduce a hybrid approach that applies greedy clustering to upper tree levels, where the number of items is large, and uses the original constrained $k$-means for deeper levels, where higher clustering precision is beneficial. This design is based on the hypothesis that coarse groupings near the top of the tree can be approximated efficiently without harming retrieval quality, while precise clustering near the leaves better preserves semantic structure. Empirically, the upper levels dominate total construction time, making this hybrid strategy particularly effective in balancing efficiency and quality.

Algorithm~\ref{alg:faiss_tree} outlines the hierarchical clustering procedure combining the greedy and constrained variants described above. Importantly, our modifications affect only the tree construction phase. SEATER’s model architecture, training objectives, and inference procedure remain unchanged.

\begin{algorithm}[t]
\DontPrintSemicolon
\LinesNumbered
\SetAlgoNoLine
\SetAlgoNoEnd
\caption{Hybrid Hierarchical Clustering Algorithm}
\label{alg:faiss_tree}
\KwIn{Item embeddings $X_{1:N}$, number of clusters $k$, greedy threshold $T$}
\KwOut{Semantic item identifiers $L_{1:N}$}

\SetKwFunction{HierarchicalClustering}{HierarchicalClustering}
\SetKwProg{Fn}{Function}{:}{}
\Fn{\HierarchicalClustering{$X$}}{
    MinSize $\leftarrow \lfloor |X| / k \rfloor$, 
    MaxSize $\leftarrow \lfloor |X| / k \rfloor + 1$\;
    
    \tcp{Choose clusters strategy based on cluster size}
    \eIf{$|X| > T$}{
    \tcp{Efficient greedy clusters for large sets}
        $\mathcal{C} \leftarrow \text{FaissCluster}(X, k, \text{MaxSize}, \text{MinSize})$\;
    }{
        \tcp{More precise constrained clustering for smaller sets}
        $\mathcal{C} \leftarrow \text{ConstrainedKMeans}(X, k, \text{MaxSize}, \text{MinSize})$\;
    }

    \tcp{Recursively partition each cluster}
    \For{$i = 0$ \textbf{to} $k - 1$}{
        \eIf{$|C_{i+1}| > k$}{
            $\text{HierarchicalClustering}(C_{i+1})$\;
        }{
             \tcp{Leaf node: assign final identifiers}
            Assign sequential identifiers to $C_{i+1}$\;
        }
    }

    Construct hierarchical identifiers $L$ based on recursive cluster indices\;
    \Return{$L$}\;
}
\end{algorithm}

\smallskip
In summary, our greedy and hybrid clustering algorithms address SEATER’s scalability bottleneck by replacing the global capacity-constrained optimization with a parallelizable local heuristic and combining it with precise clustering near the tree leaves. These methods reduce tree construction time by up to two orders of magnitude, down to less than 2\% of the original on the largest dataset, while preserving inference efficiency.

%% file: sections/sec-exp.tex
\section{Experimental setup}

In this section, we introduce the datasets used for the reproduction of SEATER, including the additional Yambda dataset. Then, we describe the baselines and implementation details of our reproduction and outline the evaluation protocol for assessing retrieval quality and efficiency.

\subsection{Datasets}

Following the original work \cite{seater}, we evaluate SEATER on Yelp, Books, and News, and extend the evaluation to a larger-scale dataset, Yambda. Table~\ref{tab:dataset_statistics} summarizes the main statistics.

\vspace{-14pt}
\subsubsection{Yelp}
Yelp contains user-business interactions and reviews.\footnote{\url{https://business.yelp.com/data/resources/open-dataset/}} We use the standard train, validation, and test splits provided in the LightGCN repository \cite{10.1145/3397271.3401063}. This setup was verified with the SEATER authors to align with their original preprocessing.

\vspace{-14pt}
\subsubsection{Books}
The Books subset of the Amazon review dataset \cite{ni-etal-2019-justifying}.\footnote{\url{https://nijianmo.github.io/amazon/}} 
We follow the ComiRec preprocessing pipeline \cite{comirec}, identical to the one used in SEATER’s original implementation.

\vspace{-14pt}
\subsubsection{News}
The small partition of the MIcrosoft News Dataset (MIND) \cite{wu-etal-2020-mind}.\footnote{\url{https://msnews.github.io/}} 
Since the original preprocessing code was unavailable, we reconstructed the setup based on author correspondence. 
We used the training data of MIND-small and split users by an 8:1:1 ratio into train, validation, and test sets. 
For each user, positive impressions were concatenated with their interaction history, and sequences were expanded into multiple training instances. 
During evaluation, we used the first 80\% of each sequence as input context and the remaining 20\% as targets, consistent with the other datasets.

\vspace{-14pt}
\subsubsection{Yambda}
Yambda \cite{ploshkin2025yambda5blargescalemultimodal} is a large-scale open dataset of user-music interactions from Yandex Music.\footnote{\url{https://huggingface.co/datasets/yandex/yambda}} 
With a substantially larger item pool, it serves as a scalability stress test, clearly exposing tree construction as a bottleneck during training and motivating our proposed algorithms. Following the dataset creators’ preprocessing, we keep only organic interactions and filter out tracks played for less than 50\% of their duration. Data is temporally split into 299 days for training, a 30-minute gap, one day for validation, and one day for testing.

\begin{table}[t!]
\centering
\caption{Dataset statistics.}
\setlength{\tabcolsep}{8pt}
\begin{tabular}{lcccc}
\toprule
\textbf{} & \textbf{Yelp} & \textbf{News} & \textbf{Books} & \textbf{Yambda} \\
\midrule
\textbf{\#Users} & 31,668 & 50,000 & 482,934 & 8,798 \\
\textbf{\#Items} & 38,048 & 39,865 & 361,245 & 475,910 \\
\textbf{\#Interactions} & 1,561,406 & 1,162,402 & 8,898,041 & 46,467,212 \\
\bottomrule
\end{tabular}
\label{tab:dataset_statistics}
\end{table}

\subsection{Reproduction details}

All experiments were run on a single NVIDIA H100 GPU. We use the official SEATER implementation\footnote{\url{https://github.com/Ethan00Si/SEATER_Generative_Retrieval}} for model architecture, training, and tree construction, keeping hyperparameters consistent across datasets. Learning rate was tuned in the range [1e-3, 1e-6], with 0.001 selected as optimal across datasets. We keep all hyperparameters aligned with the original work: We use an embedding dimension of 64, a feed-forward dimension of 256, a dropout of 0.2, and early stopping with patience between 4 and 6 epochs, depending on the dataset size.

\vspace{-16pt}

\subsubsection{Baselines}
We compare SEATER against the baselines reported in the original paper, using the authors’ published results.  To verify reproducibility and ensure consistent evaluation, we re-implement three representative baselines: GRU4Rec, SASRec, and BERT4Rec, following the configurations described in the original work \cite{seater}. GRU4Rec is implemented via the ComiRec repository.\footnote{\url{https://github.com/THUDM/ComiRec/tree/master}} We implement BERT4Rec using an open-source implementation\footnote{\url{https://github.com/jaywonchung/BERT4Rec-VAE-Pytorch}}, and SASRec based on the SEATER authors’ code. GRU4Rec and SASRec are trained autoregressively by predicting the next item from user history, while BERT4Rec uses a bidirectional Cloze-style objective. Learning rates are selected from \{0.001, 0.0001\}, with 0.001 performing best in most cases and 0.0001 used for BERT4Rec on the News dataset.

\vspace{-16pt}

\subsubsection{Tree construction}
We build SEATER’s identifier tree using SASRec embeddings trained on the same dataset. These embeddings are only used during tree construction, and then discarded. We use the tree construction implementation provided by the original authors \cite{seater}, which uses constrained $k$-means \cite{Levy-Kramer_k-means-constrained_2018}. Our greedy variant replaces it with a Faiss-based approximate nearest-neighbor search \cite{douze2024faiss}, and the hybrid method switches from greedy to constrained clustering once the number of items per level drops below 2,000, an empirically chosen threshold balancing efficiency and clustering precision. 

\vspace{-4pt}

\subsubsection{Evaluation protocol}
Following \cite{seater}, we evaluate retrieval accuracy using Recall@k, Hit Rate@k (HR@k), and Normalized Discounted Cumulative Gain@k (NDCG@k) for $k \in \{20, 50\}$, averaged across test users. To assess training efficiency, we measure total training time (excluding tree construction) and tree-construction time. This allows us to isolate the cost of tree building and highlight its impact as a potential training time bottleneck. It is worth noting that, next to our reported tree-construction and training times, the full SEATER pipeline also includes training SASRec to obtain item embeddings. Following the original work \cite{seater}, we assume that practitioners already have access to an existing embedding model and therefore exclude SASRec training time from our comparisons.

%% file: sections/sec-results.tex
\section{Results}
\subsection{RQ1: Reproduction and generalization of SEATER's results}

Our results for SEATER on Books, Yelp, and News, reported in ~\autoref{tab:performance_comparison_reproduction}, are very comparable to those from the original work. On Books and Yelp, our SEATER results closely match or show minor deviations from the original. On the News dataset, our implementation of SEATER achieves slightly higher scores than the original results.

The superior performance of SEATER over the baselines holds for Books and Yelp. However, on the News dataset, while SEATER still outperforms GRU4Rec, the performance gap is significantly narrower than in the original paper. For instance, our GRU4Rec's Recall@50 is very close to SEATER's, unlike the wider gap in the original report. We hypothesize that this stems from our baseline implementations and preprocessing for News, as specific preprocessing details and original code were unavailable.

\autoref{tab:performance_comparison_yambda} shows our results on the additional dataset, Yambda. SEATER maintains top performance across all metrics, indicating its robustness. Although differences between SEATER and some baselines are not in every setting as clear as in the original work, we answer \ref{rq1} positively: The reported performance gains of SEATER over baselines can be replicated and generalized to the additional large-scale dataset Yambda.

\begin{table}[t!]
\centering
\caption{
Comparison of models across datasets. 
\textcolor{blue!60}{\textbf{Blue}} columns indicate models reproduced in this work, 
and \textcolor{gray}{\textbf{gray}} columns indicate results reported in the original SEATER paper \cite{seater}. The best scores per metric are in \textbf{bold}.
}
\label{tab:performance_comparison_reproduction}
\begin{adjustbox}{width=\textwidth}
\footnotesize
\begin{tabular}{llcccccccc}
\toprule
 & & \multicolumn{3}{c}{\textbf{Dual-encoder}} 
   & \multicolumn{1}{c}{\textbf{Tree-based}} 
   & \multicolumn{4}{c}{\textbf{Generative}} \\
\cmidrule(lr){3-5} \cmidrule(lr){6-6} \cmidrule(lr){7-10}
\textbf{Dataset} & \textbf{Metric} 
& \cellcolor{blue!15}\textbf{GRU4Rec} 
& \cellcolor{blue!15}\textbf{BERT4Rec} 
& \cellcolor{blue!15}\textbf{SASRec} 
& \cellcolor{gray!20}\textbf{TDM} 
& \cellcolor{gray!20}\textbf{GPTRec} 
& \cellcolor{gray!20}\textbf{TIGER} 
& \cellcolor{gray!20}\shortstack{\textbf{SEATER}\\\textbf{(original)}} 
& \cellcolor{blue!15}\shortstack{\textbf{SEATER}\\\textbf{(ours)}} \\
\midrule

\multirow{6}{*}{Books}
& NDCG@20 & 0.0297 & 0.0436 & 0.0408 & 0.0235 & 0.0271 & 0.0468 & 0.0592 & \textbf{0.0593} \\
& NDCG@50 & 0.0401 & 0.0564 & 0.0536 & 0.0330 & 0.0373 & 0.0573 & 0.0713 & \textbf{0.0714} \\
& Hit@20 & 0.1275 & 0.1708 & 0.1689 & 0.1101 & 0.1181 & 0.1637 & \textbf{0.2006} & 0.2000 \\
& Hit@50 & 0.2055 & 0.2571 & 0.2578 & 0.1832 & 0.1962 & 0.2380 & \textbf{0.2813} & 0.2815 \\
& Recall@20 & 0.0573 & 0.0796 & 0.0807 & 0.0475 & 0.0533 & 0.0766 & \textbf{0.0972} & 0.0967 \\
& Recall@50 & 0.0981 & 0.1296 & 0.1311 & 0.0849 & 0.0938 & 0.1179 & \textbf{0.1448} & 0.1442 \\
\midrule

\multirow{6}{*}{Yelp}
& NDCG@20 & 0.0452 & 0.0463 & 0.0468 & 0.0414 & 0.0440 & 0.0539 & \textbf{0.0572} & 0.0527 \\
& NDCG@50 & 0.0668 & 0.0667 & 0.0694 & 0.0610 & 0.0653 & 0.0769 & \textbf{0.0810} & 0.0756 \\
& Hit@20 & 0.3584 & 0.3616 & 0.3762 & 0.3493 & 0.3487 & 0.4087 & \textbf{0.4201} & 0.3974 \\
& Hit@50 & 0.5644 & 0.5627 & 0.5859 & 0.5439 & 0.5527 & 0.5922 & \textbf{0.6118} & 0.5940 \\
& Recall@20 & 0.0563 & 0.0561 & 0.0597 & 0.0524 & 0.0559 & 0.0679 & \textbf{0.0720} & 0.0654 \\
& Recall@50 & 0.1141 & 0.1113 & 0.1195 & 0.1040 & 0.1121 & 0.1271 & \textbf{0.1353} & 0.1268 \\
\midrule

\multirow{6}{*}{News}
& NDCG@20 & 0.0930 & 0.0759 & 0.0916 & 0.0830 & 0.0813 & 0.0919 & 0.0942 & \textbf{0.0959} \\
& NDCG@50 & 0.1229 & 0.1011 & 0.1209 & 0.1067 & 0.1065 & 0.1182 & 0.1225 & \textbf{0.1251} \\
& Hit@20 & 0.4083 & 0.3731 & 0.4106 & 0.3821 & 0.3731 & 0.4019 & 0.4070 & \textbf{0.4236} \\
& Hit@50 & 0.5801 & 0.5431 & 0.5792 & 0.5248 & 0.5305 & 0.5531 & 0.5747 & \textbf{0.5810} \\
& Recall@20 & 0.1507 & 0.1312 & 0.1516 & 0.1280 & 0.1324 & 0.1408 & 0.1456 & \textbf{0.1581} \\
& Recall@50 & 0.2541 & 0.2176 & 0.2524 & 0.2080 & 0.2182 & 0.2292 & 0.2429 & \textbf{0.2581} \\
\bottomrule
\end{tabular}
\end{adjustbox}
\end{table}

\begin{table}[t!]
\centering
\caption{Performance comparison of SEATER and baselines on Yambda. The best result per metric is in \textbf{bold} and the second-best is \underline{underscored}.}
\label{tab:performance_comparison_yambda}

\begin{tabular}{lcccc}
\toprule
 & \multicolumn{3}{c}{\textbf{Dual-encoder}} & \multicolumn{1}{c}{\textbf{Generative}} \\
\cmidrule(lr){2-4} \cmidrule(lr){5-5}
\textbf{Metric} & \textbf{GRU4Rec} & \textbf{BERT4Rec} & \textbf{SASRec} & \textbf{SEATER} \\
\midrule
NDCG@20   & 0.0680 & 0.0412 & \underline{0.0960} & \textbf{0.1245} \\
NDCG@50   & 0.0820 & 0.0503 & \underline{0.1138} & \textbf{0.1370} \\
Hit@20    & 0.3504 & 0.2481 & \underline{0.4298} & \textbf{0.4566} \\
Hit@50    & 0.4745 & 0.3499 & \underline{0.5406} & \textbf{0.5485} \\
Recall@20 & 0.0701 & 0.0448 & \underline{0.1006} & \textbf{0.1175} \\
Recall@50 & 0.1203 & 0.0776 & \underline{0.1636} & \textbf{0.1705} \\
\bottomrule
\end{tabular}
\end{table}

\subsection{RQ2: The impact of greedy tree construction}

Using greedy tree construction in SEATER substantially reduces tree-building time across all datasets, as shown in \autoref{tab:dataset_model_comparison_final_times} and \autoref{fig:tree-time-method}. 
The improvement is particularly pronounced for datasets with larger item catalogs, such as Books and Yambda, where tree-construction time decreases from 54 minutes to 93 seconds and from 82 minutes to 83 seconds, respectively, representing a 35$\times$ and 59$\times$ speedup. Interestingly, even when excluding tree construction, overall training time also decreases, likely due to faster model convergence. We leave it to future work to investigate the relationship between item identifier structure and convergence dynamics. Despite the use of an approximate (greedy) clustering algorithm, the performance drop across all metrics remains minimal, with the largest observed difference being only $-0.45\%$ in Recall@20 on the Books dataset. To answer \ref{rq2}, the proposed greedy clustering method substantially improves training efficiency with a minimal drop in retrieval quality.

\subsection{RQ3: Balancing greedy and complete clustering}

\begin{figure}[!t]
    \centering
    \begin{subfigure}{0.48\linewidth}
        \centering
        \includegraphics[width=\linewidth]{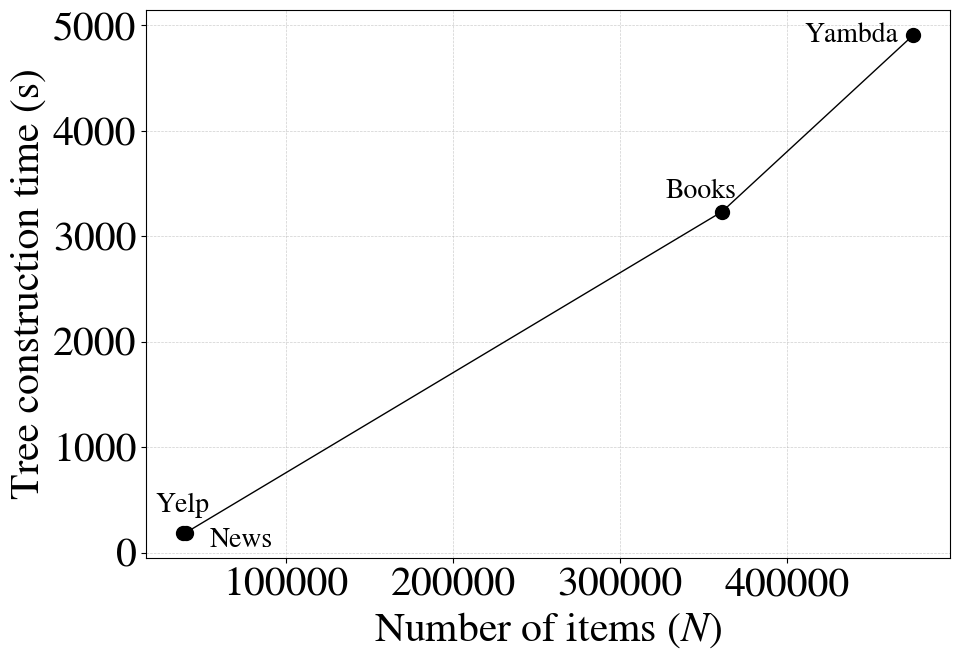}
        \caption{Tree construction time as a function of the number of items.}
        \label{fig:tree-time-items}
    \end{subfigure}
    \hfill
    \begin{subfigure}{0.48\linewidth}
        \centering
        \includegraphics[width=\linewidth]{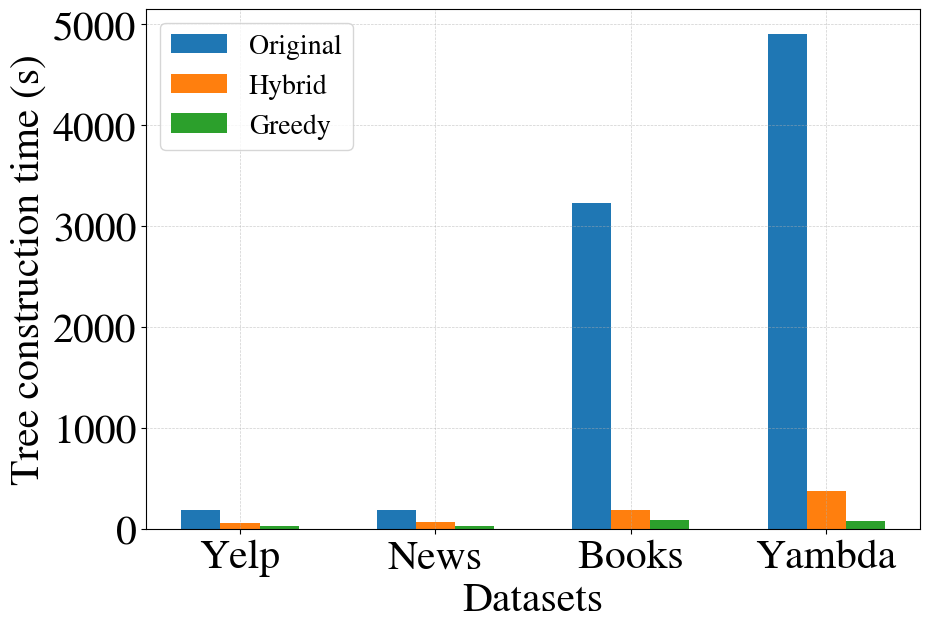}
        \caption{Tree construction time by method, for all datasets.}
        \label{fig:tree-time-method}
    \end{subfigure}
    \caption{Tree construction time in seconds. The original tree building time scales non-linearly with the number of items for the baseline algorithm, while greedy clustering and the hybrid approach lead to significantly faster tree construction, particularly for larger datasets.}
    \label{fig:tree-time}
\end{figure}

Although the performance degradation observed with greedy clustering was minimal, we propose a hybrid approach to better balance training efficiency and retrieval quality. As shown in \autoref{tab:dataset_model_comparison_final_times} and \autoref{fig:tree-time-method}, this hybrid method achieves a substantial reduction in tree-construction time compared to the original constrained $k$-means implementation (e.g., from 82 to 6 minutes on Yambda and from 54 to 3 minutes on Books), while maintaining nearly identical, or in some cases improved, retrieval performance. Interestingly, it even surpasses the original SEATER on Yambda (+2.5\% NDCG@20) and shows only marginal losses ($\leq$0.2\% NDCG@20) on Books and Yelp. To answer \ref{rq3}, our findings indicate that combining approximate and exact clustering leads to improved training efficiency and competitive retrieval quality, enabling SEATER to perform efficiently on large-scale recommendation datasets.

\begin{table}[t!]
\centering
\scriptsize
\setlength{\tabcolsep}{2.5pt} 
\renewcommand{\arraystretch}{0.95}
\caption{Comparison of tree construction algorithms across datasets and evaluation metrics. `Original' refers to the constrained k-means implementation from the original authors. The best result per metric is in \textbf{bold} and the second-best is \underline{underscored}. SASRec is included as a reference. Training time does not include tree construction time. Results are reported over 3 seeds. Training times reported as (h:m:s).}
\label{tab:dataset_model_comparison_final_times}
\begin{tabular}{@{}llcccc@{}}
\toprule
 & & \multicolumn{1}{c}{\textbf{Dual-encoder}} & \multicolumn{3}{c}{\textbf{SEATER}} \\
\cmidrule(lr){3-3} \cmidrule(lr){4-6}
\textbf{Dataset} & \textbf{Metric} & \textbf{SASRec} & \textbf{Original} & \textbf{Hybrid} & \textbf{Greedy} \\
\midrule

\multirow{8}{*}{\textbf{News}}
& NDCG@20   & 0.0916$\pm$0.9{e}{-3}  & \textbf{0.0959$\pm$1.0{e}{-3}} & 0.0950$\pm$1.0{e}{-3} & \underline{0.0957$\pm$2.4{e}{-3}} \\
& NDCG@50   & 0.1209$\pm$0.8{e}{-3}  & \textbf{0.1251$\pm$1.2{e}{-3}} & 0.1246$\pm$1.1{e}{-3} & \underline{0.1249$\pm$2.4{e}{-3}} \\
& Recall@20 & 0.1516$\pm$2.2{e}{-3}  & \textbf{0.1581$\pm$1.5{e}{-3}} & 0.1548$\pm$1.6{e}{-3} & \underline{0.1552$\pm$3.3{e}{-3}} \\
& Recall@50 & 0.2524$\pm$2.9{e}{-3}  & \textbf{0.2581$\pm$3.1{e}{-3}} & \underline{0.2576$\pm$1.6{e}{-3}} & 0.2555$\pm$2.4{e}{-3} \\
& Hit@20    & 0.4106$\pm$8.6{e}{-3}  & \textbf{0.4236$\pm$2.6{e}{-3}} & 0.4151$\pm$5.0{e}{-3} & \underline{0.4155$\pm$8.2{e}{-3}} \\
& Hit@50    & 0.5792$\pm$7.3{e}{-3}  & \underline{0.5810$\pm$1.1{e}{-3}} & \textbf{0.5812$\pm$0.4{e}{-3}} & 0.5755$\pm$7.4{e}{-3} \\
& Tree time (s) & -- & 190.02 $\pm$ 2.55 & \underline{67.08 $\pm$ 0.71} & \textbf{31.61 $\pm$ 0.72} \\
& Training time & 0:11:13 & 1:37:48 & 1:37:18 & 1:35:56 \\
\midrule

\multirow{8}{*}{\textbf{Yelp}}
& NDCG@20   & 0.0468$\pm$1.8{e}{-3}  & \textbf{0.0527$\pm$2.0{e}{-3}} & \underline{0.0515$\pm$0.6{e}{-3}} & 0.0500$\pm$1.0{e}{-3} \\
& NDCG@50   & 0.0694$\pm$0.4{e}{-3}  & \textbf{0.0756$\pm$1.6{e}{-3}} & \underline{0.0742$\pm$0.6{e}{-3}} & 0.0720$\pm$2.3{e}{-3} \\
& Recall@20 & 0.0597$\pm$1.8{e}{-3}  & \textbf{0.0654$\pm$2.1{e}{-3}} & \underline{0.0639$\pm$0.7{e}{-3}} & 0.0616$\pm$1.1{e}{-3} \\
& Recall@50 & 0.1195$\pm$1.6{e}{-3}  & \textbf{0.1268$\pm$1.6{e}{-3}} & \underline{0.1249$\pm$0.7{e}{-3}} & 0.1208$\pm$4.8{e}{-3} \\
& Hit@20    & 0.3762$\pm$8.8{e}{-3}  & \textbf{0.3974$\pm$9.8{e}{-3}} & \underline{0.3966$\pm$9.8{e}{-3}} & 0.3844$\pm$7.1{e}{-3} \\
& Hit@50    & 0.5859$\pm$3.3{e}{-3}  & \textbf{0.5940$\pm$1.3{e}{-3}} & \underline{0.5900$\pm$5.3{e}{-3}} & 0.5791$\pm$8.7{e}{-3} \\
& Tree time (s) & -- & 189.35 $\pm$ 3.61 & \underline{61.27 $\pm$ 0.23} & \textbf{31.15 $\pm$ 0.34} \\
& Training time  & 0:18:42 & 1:10:33 & 1:10:14 & 0:59:42 \\
\midrule

\multirow{8}{*}{\textbf{Books}}
& NDCG@20   & 0.0408$\pm$0.4{e}{-3}  & \textbf{0.0593$\pm$0.1{e}{-3}} & \underline{0.0574$\pm$0.5{e}{-3}} & 0.0566$\pm$0.3{e}{-3} \\
& NDCG@50   & 0.0536$\pm$0.5{e}{-3}  & \textbf{0.0714$\pm$0.2{e}{-3}} & \underline{0.0691$\pm$0.5{e}{-3}} & 0.0682$\pm$0.4{e}{-3} \\
& Recall@20 & 0.0807$\pm$0.7{e}{-3}  & \textbf{0.0967$\pm$0.3{e}{-3}} & \underline{0.0936$\pm$0.5{e}{-3}} & 0.0922$\pm$0.4{e}{-3} \\
& Recall@50 & 0.1311$\pm$1.3{e}{-3}  & \textbf{0.1442$\pm$0.3{e}{-3}} & \underline{0.1398$\pm$0.3{e}{-3}} & 0.1374$\pm$0.3{e}{-3} \\
& Hit@20    & 0.1689$\pm$1.2{e}{-3}  & \textbf{0.2000$\pm$0.8{e}{-3}} & \underline{0.1960$\pm$0.9{e}{-3}} & 0.1931$\pm$0.3{e}{-3} \\
& Hit@50    & 0.2578$\pm$2.0{e}{-3}  & \textbf{0.2815$\pm$1.5{e}{-3}} & \underline{0.2756$\pm$0.3{e}{-3}} & 0.2720$\pm$0.2{e}{-3} \\
& Tree time (s) & -- & 3232.80 $\pm$ 14.09 & \underline{184.05 $\pm$ 0.56} & \textbf{93.32 $\pm$ 3.92} \\
& Training time  & 0:26:49 & 3:54:27 & 3:04:38 & 2:28:32 \\
\midrule

\multirow{8}{*}{\textbf{Yambda}}
& NDCG@20   & 0.0960$\pm$17.8{e}{-3} & 0.1245$\pm$7.3{e}{-3} & \textbf{0.1495$\pm$21.0{e}{-3}} & \underline{0.1399$\pm$17.9{e}{-3}} \\
& NDCG@50   & 0.1138$\pm$21.0{e}{-3} & 0.1370$\pm$8.0{e}{-3} & \textbf{0.1601$\pm$20.4{e}{-3}} & \underline{0.1493$\pm$17.6{e}{-3}} \\
& Recall@20 & 0.1006$\pm$20.0{e}{-3} & 0.1175$\pm$8.0{e}{-3} & \textbf{0.1461$\pm$23.6{e}{-3}} & \underline{0.1353$\pm$20.1{e}{-3}} \\
& Recall@50 & 0.1636$\pm$29.9{e}{-3} & 0.1705$\pm$11.7{e}{-3} & \textbf{0.1966$\pm$24.0{e}{-3}} & \underline{0.1817$\pm$22.1{e}{-3}} \\
& Hit@20    & 0.4298$\pm$34.2{e}{-3} & 0.4566$\pm$18.0{e}{-3} & \textbf{0.4975$\pm$28.3{e}{-3}} & \underline{0.4841$\pm$27.1{e}{-3}} \\
& Hit@50    & 0.5406$\pm$35.3{e}{-3} & 0.5485$\pm$17.6{e}{-3} & \textbf{0.5795$\pm$24.5{e}{-3}} & \underline{0.5638$\pm$27.2{e}{-3}} \\
& Tree time (s) & -- & 4907.97 $\pm$ 68.16 & \underline{375.72 $\pm$ 3.31} & \textbf{83.21 $\pm$ 2.14} \\
& Training time  & 0:47:16 & 5:39:04 & 4:23:58 & 2:37:05 \\
\bottomrule
\end{tabular}
\end{table}

\vspace{-6pt}
\subsection{Ablation Study}
To further validate our method, we perform an ablation over the switch depth at which we transition from clustering-based construction to greedy matching. On Books (Table \ref{tab:books_vertical_decimal}), increasing the switch depth improves ranking quality but substantially increases tree construction time, revealing a clear efficiency-effectiveness tradeoff. The fastest construction occurs at depth 1000, while the strongest ranking performance is achieved at depth 50000, at the cost of an order-of-magnitude slower build time. Importantly, gains exhibit diminishing returns: depth 2000 achieves performance comparable to depth 50000 (e.g., nDCG@20 0.0572 vs 0.0585) while reducing construction time substantially (187.07s vs 1452.15s).

This pattern is even more pronounced on Yambda (Table \ref{tab:yambda_vertical_decimal}), where depth 2000 provides the best ranking metrics overall, while deeper thresholds incur large additional construction costs with little or no additional benefit. Overall, depth 2000 appears to be a robust operating point that balances ranking quality and construction efficiency across datasets.

\begin{table}[h!]
\centering
\caption{Performance on the \textbf{Yambda} dataset across different thresholds. Results show mean $\pm$ standard deviation over 3 seeds.}
\label{tab:yambda_vertical_decimal}
\scriptsize
\setlength{\tabcolsep}{4pt} 
\begin{tabular}{@{}lcccc@{}}
\toprule
& \multicolumn{4}{c}{\textbf{Threshold}} \\
\cmidrule(lr){2-5}
\textbf{Metric} & \textbf{100} & \textbf{1000} & \textbf{2000} & \textbf{50000} \\
\midrule
Time (s) & \textbf{79.95 $\pm$ 3.28} & \underline{129.99 $\pm$ 2.40} & 380.35 $\pm$ 4.80 & 2476.85 $\pm$ 14.05 \\
\midrule
nDCG@20 & 0.1507 $\pm$ 0.0169 & 0.1495 $\pm$ 0.0183 & \textbf{0.1616 $\pm$ 0.0009} & \underline{0.1509 $\pm$ 0.0163} \\
nDCG@50 & 0.1600 $\pm$ 0.0168 & 0.1591 $\pm$ 0.0173 & \textbf{0.1724 $\pm$ 0.0009} & \underline{0.1617 $\pm$ 0.0161} \\
Hit@20 & \underline{0.4994 $\pm$ 0.0246} & 0.4956 $\pm$ 0.0268 & \textbf{0.5133 $\pm$ 0.0015} & 0.4986 $\pm$ 0.0250 \\
Hit@50 & 0.5751 $\pm$ 0.0223 & 0.5721 $\pm$ 0.0215 & \textbf{0.5932 $\pm$ 0.0038} & \underline{0.5788 $\pm$ 0.0234} \\
Recall@20 & 0.1462 $\pm$ 0.0178 & 0.1449 $\pm$ 0.0206 & \textbf{0.1589 $\pm$ 0.0009} & \underline{0.1475 $\pm$ 0.0182} \\
Recall@50 & 0.1930 $\pm$ 0.0200 & 0.1928 $\pm$ 0.0201 & \textbf{0.2110 $\pm$ 0.0014} & \underline{0.1981 $\pm$ 0.0199} \\
\bottomrule
\end{tabular}
\end{table}

\begin{table}[t!]
\centering
\caption{Performance on the \textbf{Books} dataset across different thresholds. Results show mean $\pm$ standard deviation over 3 seeds.}
\label{tab:books_vertical_decimal}
\scriptsize
\setlength{\tabcolsep}{4pt} 
\begin{tabular}{@{}lcccc@{}}
\toprule
& \multicolumn{4}{c}{\textbf{Threshold}} \\
\cmidrule(lr){2-5}
\textbf{Metric} & \textbf{50} & \textbf{1000} & \textbf{2000} & \textbf{50000} \\
\midrule
Time (s) & \underline{91.52 $\pm$ 0.46} & \textbf{90.22 $\pm$ 0.61} & 187.07 $\pm$ 1.49 & 1452.48 $\pm$ 18.28 \\
\midrule
nDCG@20 & 0.0566 $\pm$ 0.0003 & 0.0562 $\pm$ 0.0009 & \underline{0.0572 $\pm$ 0.0007} & \textbf{0.0585 $\pm$ 0.0002} \\
nDCG@50 & 0.0682 $\pm$ 0.0003 & 0.0678 $\pm$ 0.0009 & \underline{0.0689 $\pm$ 0.0006} & \textbf{0.0703 $\pm$ 0.0003} \\
Hit@20 & 0.1931 $\pm$ 0.0002 & 0.1922 $\pm$ 0.0027 & \underline{0.1958 $\pm$ 0.0009} & \textbf{0.1974 $\pm$ 0.0013} \\
Hit@50 & 0.2720 $\pm$ 0.0001 & 0.2713 $\pm$ 0.0020 & \underline{0.2754 $\pm$ 0.0005} & \textbf{0.2774 $\pm$ 0.0012} \\
Recall@20 & 0.0922 $\pm$ 0.0003 & 0.0917 $\pm$ 0.0013 & \underline{0.0938 $\pm$ 0.0005} & \textbf{0.0949 $\pm$ 0.0004} \\
Recall@50 & 0.1374 $\pm$ 0.0002 & 0.1368 $\pm$ 0.0013 & \underline{0.1394 $\pm$ 0.0002} & \textbf{0.1413 $\pm$ 0.0009} \\
\bottomrule
\end{tabular}
\end{table}
\vspace{-12pt}

%% file: sections/sec-con.tex
\section{Conclusion}

We successfully reproduce the core findings of SEATER, validating its increased retrieval performance due to its semantic tree-structured identifiers and training strategy. Our experiments verify SEATER’s effectiveness across both the original datasets and the newly introduced Yambda benchmark, where it maintained strong retrieval quality. 

To overcome the training bottleneck of item identifier tree construction, we proposed a greedy clustering method, which substantially reduces tree construction time by up to 59×, with only a minor drop in retrieval quality. A hybrid method, which combines greedy clustering at higher tree levels with constrained $k$-means at deeper levels, achieves a trade-off between training efficiency and retrieval quality by reducing tree construction time with minimal impact on final performance. These findings enable SEATER to scale to larger datasets and settings that require regular retraining of models, without sacrificing its retrieval quality and inference efficiency.

\vspace{-10pt}

\subsection{Limitations and Future Work}

While our study confirms the effectiveness and scalability of SEATER, a few limitations remain. 
In line with the original work, we exclusively use item embeddings derived from SASRec. Future work could explore alternative or hybrid embedding strategies, such as incorporating item metadata (e.g., genres, descriptions) alongside collaborative filtering signals, to better capture semantic item relationships. Additionally, our improved tree construction time could allow for the exploration of alternative methods to create and adapt the identifier tree, such as rebuilding the tree during training time to better capture hierarchy. In our experiments, we observed a change in model convergence depending on the tree-construction algorithm. We leave it to future work to explore this relationship, possibly leading to further improvements in training time through novel item identifier designs.

\vspace{-10pt}
\subsubsection{Acknowledgments}
This research was (partially) supported by the Dutch Research Council (NWO), under project number KICH3.LTP.20.006.

\vspace{-10pt}
\subsubsection{Disclosure of Interests}
The authors have no competing interests to declare that are relevant to the content of this article. 